\documentclass[12pt,amsmath,amssymb]{article}
\usepackage{epsfig,amsmath,amssymb,axodraw}
\usepackage{graphicx}% Include figure files
\usepackage{dcolumn}% Align table columns on decimal point
\usepackage{bm}% bold math
\usepackage{hyperref}

\newcommand{\be}{\begin{equation} } 
\newcommand{\ee}{\end{equation} } 
\newcommand{\ba}{\begin{array} } 
\newcommand{\ea}{\end{array} } 
\newcommand{\bear}{\begin{eqnarray} } 
\newcommand{\eear}{\end{eqnarray} } 
\newcommand{\met}{ E_T\!\!\! \!\! \! \slash \;\, } 

\begin{document}

\baselineskip=21pt \pagestyle{plain} \setcounter{page}{1}

\vspace*{-1.1cm}

\begin{flushright}{\small Fermilab-PUB-15-259-T}\end{flushright}

\vspace*{0.2cm}

\begin{center}

{\large \bf Leptophobic boson signals with leptons, jets and missing energy}\\ [9mm]

{\normalsize \bf Bogdan A. Dobrescu   
 \\ [4mm]
{\it
Theoretical Physics Department, Fermilab, Batavia, IL 60510, USA \\ [2mm]
}
}

\vspace*{0.5cm}

June 14, 2015

\vspace*{1.7cm}

{\bf \small Abstract}

\vspace*{-0.2cm}

\end{center}
Color-singlet gauge bosons with renormalizable couplings to quarks but not to leptons must interact with 
additional fermions (``anomalons") required to cancel the gauge anomalies.
Analyzing the decays of such leptophobic bosons into anomalons, I show that they produce
final states involving leptons at the LHC.
Resonant production of a flavor-universal leptophobic $Z'$ boson  leads to 
cascade decays via anomalons, whose signatures include a leptonically decaying $Z$, missing  energy and several jets. 
A $Z'$ boson that couples to the right-handed quarks of the first and second generations undergoes cascade decays
that violate lepton universality and include 
signals with two leptons and jets,  or with a Higgs boson, a lepton, a $W$ and missing energy.

\newpage

\tableofcontents

\nopagebreak
%%%%%%%%%%%%%%%%%%%%%%%%%%%%%%
\section{Introduction}

Any spin-1 particle that couples to light quarks can be produced resonantly at hadron colliders of center-of-mass energy ($\sqrt{s} \,$)
larger than its mass. Once produced, the spin-1 particle may decay into a quark and an antiquark, which hadronize into a pair of jets.
Despite the resonantly-enhanced cross section, the searches for dijet peaks \cite{Aad:2014aqa} set  relatively weak \cite{Dobrescu:2013cmh}  
limits on the coupling of new heavy bosons to quarks, due to the large QCD background. 
If the spin-1 particle can also decay into a lepton pair, then the coupling limits are much more stringent due to the dilepton resonance searches \cite{Agashe:2014kda}. 
The latter, however, are irrelevant in the case of leptophobic bosons, {\it i.e.} those with vanishing couplings to leptons \cite{Babu:1996vt,Rosner:1996eb,Carena:2004xs}. 

Theories that contain leptophobic bosons, however, must include new fermions or scalars, implying that the 
spin-1 particle may decay into final states with more tractable backgrounds than two jets.
The argument is based on the self-consistency of theories that include spin-1 particles. The UV behavior of vector bosons requires that either these are composite 
particles, or else they are associated with a new gauge symmetry. The former case would lead to a plethora of new states with a potentially rich phenomenology, but will not be analyzed here.

In the case of an extension of the Standard Model (SM) gauge symmetry, a few extra conditions must be satisfied. The new gauge symmetry must be 
spontaneously broken, so that there is an additional sector that includes new 
particles. The gauge boson may decay into these, as shown in \cite{Georgi:1996ei} for the case of a $Z'$ boson, in \cite{Dobrescu:2013gza} 
for a $W'$ boson, and  in \cite{Bai:2010dj}  for a color-octet spin-1 particle (an intrinsically leptophobic boson). 

Furthermore, the gauge symmetry must be free of anomalies \cite{Bardeen:1969md}. 
The SM quarks are chiral fermions, so if they are charged under the new symmetry\footnote{It is possible that the 
SM quarks are not charged under the new symmetry and yet couple to the new gauge boson, through mixing with 
some vectorlike quarks charged under the new gauge group \cite{Fox:2011qd}.}, then there are also mixed gauge anomalies that need to be canceled.
In addition, to avoid large flavor-changing neutral currents induced by gauge boson exchange, the left-handed quark doublets must carry the same 
charges.
It turns out that any color-singlet\footnote{A heavy color-octet gauge boson 
does not require new fermions \cite{Bai:2010dj}; interestingly, its dijet resonant signal may be experimentally distinguished from a color-singlet \cite{Chivukula:2014pma}.} 
leptophobic gauge boson whose couplings to quark doublets are generation-independent
requires new fermions to cancel the anomalies \cite{Dobrescu:2014fca}. Such fermions are usually referred to as ``anomalons".
The severe LHC constraints on additional chiral fermions imply that the anomalons must be vectorlike with respect to the SM gauge group and chiral with respect to the new gauge symmetry.

Here I study the LHC signals arising from decays of leptophobic $Z'$ bosons into anomalons. I first consider a 
$U(1)$ gauge symmetry with SM fermion charges proportional to the baryon number \cite{Carena:2004xs,Dobrescu:2013cmh}.
The gauge boson associated with this symmetry is usually labelled $Z'_B$.
The simplest anomalon content is given in \cite{Dobrescu:2014fca} and consists of three color-singlet fermion representations.
Heavier anomalons decay into a SM boson and a lighter anomalon, with the lightest of them being electrically-neutral and stable.
Thus, the $Z'_B$ gives rise to several patterns of cascade decays that involve $W$, $Z$ or Higgs bosons and missing energy.

Then I derive a simple set of color-singlet anomalons for a $U(1)_z$ gauge symmetry with the SM fermions carrying its charge being 
only the right-handed quarks of the first and second generation. The gauge boson associated with this symmetry, denoted here by $Z'_{R12}$,
has cascade decays that involve electrically-charged anomalons and lead to final states that violate lepton universality.

Although these are only two examples of the  leptophobic $Z'$ models one could imagine, their 
implications for the LHC are sufficiently rich (given their relatively simple structure)  to warrant special attention.
The observation that leptophobic $Z'$ bosons may decay into anomalons has been made previously by Rosner  \cite{Rosner:1996eb},
for a flavor-independent $Z'$ model based on the $E_6$ unified group. 
Viable sets of anomalons also exist for other flavor-dependent sets of $U(1)_z$ quark 
charges, {\it e.g.}, the $U(1)_{ds}$ model of Ref.~\cite{Dobrescu:2014fca}. 
Nevertheless, models of these type are highly restricted by FCNC constraints, the generation of SM quark masses,  and especially the gauge anomalies.
Given that the ratios of gauge charges are expected to be rational numbers, it is nontrivial to find solutions to 
the cubic and quadratic equations required to cancel the anomalies of an additional $U(1)$ gauge symmetry.

The $s$-channel production cross sections of these leptophobic $Z'$ bosons are computed in Section 2.
Implications of the $Z'_B$ cascade decays for searches at the LHC are discussed  in Section 3. 
The anomalon content  for the $Z'_{R12}$ boson and the ensuing LHC phenomenology are presented in Section 4.
The conclusions are collected in Section 5. \\

%%%%%%%%%%%%%%%%%%%%%%%%%%%%%%%%%%%%%%%%%%%%%%%%%%%%%%%%%%%%%%%%%%%%
%%%%%%%%%%%%%%%%%%%%%%%%%%%%%%%%%%%%%%%%%%%%%%%%%%%%%%%%%%%%%%%%%%%%%%%%%%%
\section{Leptophobic $U(1)$ models}  
\setcounter{equation}{0}

New $U(1)$ gauge groups, labelled generically by $U(1)_z$, 
whose charges vanish for leptons but not for quarks can be anomaly-free with the SM fermion content only if the 
 $SU(2)_W$-doublet quarks have generation-dependent charges. Thus, to avoid large FCNCs, 
leptophobic $Z'$ bosons at the TeV scale require new fermions that are chiral with respect to $U(1)_z$ 
such that the gauge anomalies cancel.
The new fermions are usually referred to as anomalons. 

Various constraints based on LHC measurements rule out new 
fermions which are chiral with respect to the SM gauge groups \cite{Dobrescu:2014fca}, so that the anomalons must belong to vectorlike representations
of $SU(3)_c\times SU(2)_W \times U(1)_Y$ (they are ``SM-vectorlike"). 
In other words, for each anomalon $\psi_L$ of  charge $z_{\psi_L}$ under $U(1)_z$ 
there is a $\psi_R$ anomalon
that has the same SM charges as $\psi_L$ but carries $U(1)_z$ charge $z_{\psi_R} \neq z_{\psi_L}$.
The couplings of the $Z'$ to fermions are given by
\be
\frac{g_z}{2} Z^\prime_\mu  \left(  J_q^\mu + J_\psi^\mu \right) ~~,
\ee
where $g_z$ is the gauge coupling, the SM quark current is $J_q^\mu$,
and the anomalon current is 
\be
J_\psi^\mu =  \sum_\psi  \left( z_{\psi_L} \bar \psi_L \gamma^\mu  \psi_L + z_{\psi_R} \bar \psi_R \gamma^\mu  \psi_R\right)  ~~.
\ee
Two sets of quark charges under $U(1)_z$ are analyzed in the remainder of this section.

%%%%%%%%%%%%%%
\begin{figure*}[t]
\begin{center}
\includegraphics[width=0.78\textwidth, angle=0]{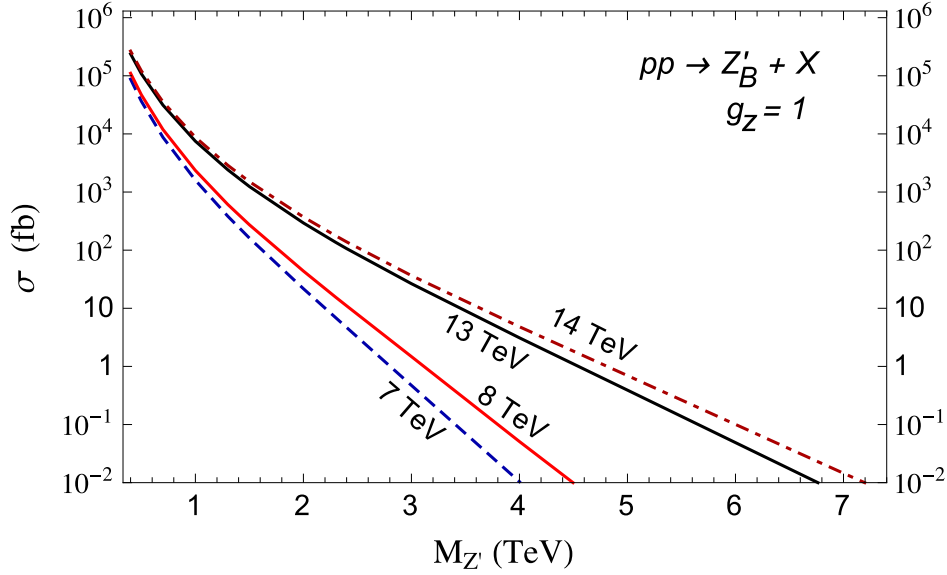} 
\caption{Cross section for $s$-channel production of $Z'_B$ at the LHC with $\sqrt{s}=7,8,13,14$ TeV,
computed at leading order for $g_z = 1$. The cross section scales as $g_z^2$ if  the 
gauge coupling satisfies $g_z \lesssim O(2)$ (at larger couplings the narrow-width approximation breaks down).
}
\label{fig:xsecZp}
\end{center}
\end{figure*}
%%%%%%%%%%%

%%%%%%%%%%%%%%
\subsection{Baryonic $Z'$}

Let us first concentrate on the case where all left- and right-handed SM quarks carry the same charge (chosen to be 1/3 without loss of generality):
\be
J_q^\mu = \frac{1}{3} \sum_q  \bar q \gamma^\mu  q  ~~,
\ee
This has the simple features that there are no tree-level FCNCs, and the quark and lepton masses are generated by the Yukawa 
couplings to the Higgs doublet, as in the SM. The gauge boson in this case, $Z'_B$, is loosely referred to as ``baryonic" $Z'$, and the 
gauge group is labelled by $U(1)_B$.

The coupling to light SM quarks allows the $s$-channel process $q \bar q \to Z'_B$, 
giving  the production cross sections at the LHC  shown in Figure \ref{fig:xsecZp} for $g_z = 1$.
The computation was performed at leading order by MadGraph5\_aMC@NLO \cite{Alwall:2014hca} (version v2\_2\_3)
using  model files generated with FeynRules \cite{Alloul:2013bka} (version 2.0.5), and 
the MSTWnlo2008 \cite{Martin:2009iq} parton distribution functions. 
Higher order corrections are expected to increase the total $Z'$ production cross section by about 20--30\% \cite{Accomando:2010fz}.
For different values of the gauge coupling $g_z$, the cross sections shown in  Figure~\ref{fig:xsecZp} scale as $g_z^2$.

The tree-level decay widths into SM quarks are 
\be
\Gamma\!\left(Z'_B \to t \bar t  \right) \simeq \frac{1}{5} \Gamma\left(Z'_B \to j j  \right) = \frac{g_z^2}{144\pi} M_{Z'}  ~~,
\ee
where $j$ stands for any hadronic jet, and $M_Z' \gg 2 m_t$ was assumed for simplicity.
Additional decays involving anomalons are possible, so that the  ratio of the total $Z'_B$ width, $\Gamma_{Z'}$,  to the $Z'_B$ mass satisfies
\be
\frac{ \Gamma_{Z'} }{M_{Z'} } = \frac{g_z^2}{24\pi \, (1- B_\psi)} ~~,
\ee
where $B_\psi$ is the sum of all branching fractions for  $Z'_B$
decays into anomalons, $B_\psi \equiv \sum_\psi B(Z'\to \psi\bar\psi)$.
The computation of the  $q \bar q \to Z'$ cross section (see Figure~1) is useful only in the narrow width approximation, {\it i.e.},
when the $\Gamma_{Z'}/M_{Z'}$ ratio is bellow $O(5\%)$. This is the case for $g_z \lesssim 2$ if 
$B_\psi \ll 1$. For larger $B_\psi$
the upper limit on $g_z$ decreases as $(1- B_\psi)^{-1/2}$; {\it e.g.},  $g_z \lesssim 0.9$ for 
 $B_\psi = 80\%$.

Several sets of $U(1)_B$ anomalons that are SM-vectorlike
have been previously identified \cite{Carena:2004xs,Dobrescu:2013cmh,Dobrescu:2014fca}.
The simplest of them is analyzed in Section 3.

\medskip

%%%%%%%%%%%%%%
\subsection{$Z'_{R12}$ boson}

$U(1)_z$ gauge groups with flavor non-universal charges for the SM quarks must satisfy a few theoretical and phenomenological constraints.
If the $Z'$ mass is of order 1 TeV, and the gauge coupling is of order one, then the weak-doublet quarks must have the same $U(1)_z$ charge 
in order to avoid large contributions to $K-\bar K$, $D - \bar D$ and $B - \bar B$ meson mixings.\footnote{In practice a small breaking 
of this $U(3)_L$ global symmetry is phenomenologically allowed for the third generation because the off-diagonal CKM elements 
that involve the $b$ quark are small.} 
The large top quark mass suggests that the $U(1)_z$  charges of the $t_R$ and $Q_L^3$ quarks are equal.

The $U(1)_z$ charges of the weak-singlet quarks may be flavor-dependent given that their gauge and mass eigenstates may be identical.
It would nevertheless be surprising if the Higgs Yukawa couplings were aligned such that both the up- and down-type singlet quarks 
are from the beginning in the mass eigenstate basis. Let's consider then the case where the gauge and mass eigenstates coincide only for 
the down-type singlet quarks ($d_R^j$ with the generations labelled by $j= 1,2,3$). Given that the FCNC processes involving the top quark 
are not severely constrained, it is sufficient to impose a $U(2)_R$ global symmetry acting on the up-type singlet quarks of the first and second generations 
($u_R^1$ and $u_R^2$). Let us normalize the gauge coupling $g_z$ such that 
$u_R^1$ and $u_R^2$ have $U(1)_z$ charges +1.

It is convenient to cancel the $[SU(3)_c]^2 U(1)_z$ gauge anomaly among the SM quarks, by assigning charge $-1$ to the $d_R^1$ and $d_R^2$ quarks.
I will label the  $Z'$ boson and the $U(1)_z$ gauge group in this model by $Z'_{R12}$ and $U(1)_{R12}$, respectively.
If the left-handed quarks and the third-generation right-handed quarks are neutral under $U(1)_{R12}$, then the masses of the first and second generation quarks 
arise from dimension-5 operators of the type $(\phi/\Lambda) \bar u_R^2  H Q_L^2$, where $\Lambda$ is the mass of a field that has been integrated out, and
$\phi$ is the scalar whose VEV breaks  $U(1)_{R12}$.
A simple set of anomalons that cancels the remaining anomalies is given in Section 4.

Given that the left- and right-handed quark currents contribute identically to the squared matrix elements for spin-1 production,
the cross section for $Z'_{R12}$ production at the LHC is larger by a factor of 9/2 than that shown in Figure 1 for $Z'_B$ (for equal $g_z$).
The partial width for $Z'_{R12}$  decay into SM quarks is 
\be
\Gamma\!\left(Z'_{R12} \to j j  \right) = \frac{g_z^2}{8\pi} M_{Z'}  ~~.
\ee
For the same branching fraction into anomalons, $B_\psi$, the mass-to-width ratio for $Z'_{R12}$ is 3 times larger than for $Z'_B$, and 
the upper limit on $g_z$ that satisfies the narrow width approximation is smaller by $\sqrt{3}$ for  $Z'_{R12}$. Thus, the upper limit 
on the  production rate  is higher for $Z'_{R12}$ than for $Z'_B$ by a factor of 3/2 (for equal $B_\psi$).

\bigskip

%%%%%%%%%%%%%%%%%%%%%%%%%%%%%%%%%%%%%%%%%%%%%%%%%%%%%%%%%%%%%%%%%%%%%%%%%%%
%%%%%%%%%%%%%%%%%%%%%%%%%%%%%%%%%%%%%%%%%%%%%%%%%%%%%%%%%%%%%%%%%%%%%%%%%%%
\section{Minimal $U(1)_B$ model}
\setcounter{equation}{0}

The minimal set of $U(1)_B$ anomalons \cite{Dobrescu:2014fca}
consists of three color-singlet SM-vectorlike fermions:
a weak doublet $L=(L^\nu,L^e)$, and two weak singlets $E$ and $N$. Their hypercharges and $U(1)_B$ charges are shown in Table 1.

The $Z'_B$ can decay into pairs of anomalons whose masses are below $M_{Z'}/2$.
The decay widths are given by
\be
\Gamma\left(Z' \to N \bar N \right) = \frac{5g_z^2}{96\pi} M_{Z'} \left( 1 - \frac{4M_N^2}{M_{Z'}^2} \right)^{\! 1/2} ~~,
\ee
and similar expressions for the other anomalons.
Due to the larger $U(1)_B$ charges, the decays into anomalons dominate over those into SM states as long as $M_N$ is not very close to $M_{Z'}/2$.
If all four anomalons are much lighter than $M_{Z'}/2$, then the total branching fraction into anomalons is $B_\psi = 5/6$.

%%%%%%%%%%%
\begin{table}[t!]\begin{center}
\renewcommand{\arraystretch}{1.2}
\begin{tabular}{cc|ccc|c}\hline 
field  & spin & $SU(3)_c$   & $SU(2)_W$   &  $U(1)_Y$  & $U(1)_B$ \\ \hline\hline
$L_L$ & & &        &                   &  \hspace*{2mm}   $\!-1$  \hspace*{2mm}  \\  [-.6em]
             & 1/2     & \ 1 \  & \  2 \ & \ $\!-1/2$  &                                                     \\  [-.6em]
$L_R$ & & &       &                   & \hspace*{2mm}   $\! +2$   \hspace*{2mm} \\ [0.1em]     
\hline
$E_L$ & & &        &                   &  \hspace*{2mm}  $\! +2$   \hspace*{2mm}  \\ [-.6em]
              & 1/2    & \ 1 \   & \  1 \ & \  $-1$  &    \\ [-.6em]
$E_R$ &  &  &     &                     &  \hspace*{2mm}  $\! -1$  \hspace*{2mm}  \\ [0.1em]
 \hline
$N_L$ & & &       &                   &   $\! +2$   
 \\ [-.6em]
          & 1/2      & \ 1 \  & \  1 \ & \  $0$ \ &    \\ [-.6em]
$N_R$ & &  &      &                   &  $\! -1$  
\\ [0.1em]      
  \hline
$\phi$ & 0 & \ 1 \  & \  1 \ & \  0 \  &   \hspace*{2mm}  $\!  +3$  \hspace*{2mm}  \\ \hline
\end{tabular}
\medskip \\
\vspace{0.2in}
\caption{New fields carrying $U(1)_B$ charge in the minimal anomalon model.}
\label{table:U1B}
\end{center}
\end{table}
%%%%%%%%%%%

The scalar field $\phi$ whose VEV is responsible for the $Z'_B$ mass must have $U(1)_B$ charge $z_\phi = 3$ so that the 
anomalons can acquire masses from the following Yukawa couplings: 
\be
- y_{E} \phi  \overline E_L E_R  -  y_{N} \phi  \overline N_L N_R - y_{L} \phi  \overline L_R L_L 
+ {\rm H. c.} 
\ee
The $y_{E}, y_{N}, y_{L}$ parameters are real and positive upon an appropriate field redefinition, for example of the left-handed anomalons.
There are also Yukawa couplings of anomalons to the Higgs doublet,
\be
- y_{EL} \overline E_L   \widetilde H  L_R - y_{LE} e^{i \theta_E}  \overline L_L H E_R
 - y_{NL}   \overline N_L H L_R - y_{LN} e^{i \theta_N}  \overline L_L  \widetilde H N_R + {\rm H. c.} 
 \label{Higgs-Yukawa}
\ee
where $\widetilde H \equiv i \sigma_2 H^*$.
The $y_{EL}$, $y_{LE}$, $y_{NL}$, $y_{LN}$ parameters can be chosen real and positive, but the two complex phases, $\theta_N$ and $\theta_E$, 
cannot be rotated away without reintroducing phases in other Yukawa couplings.

%%%%%%%%%%%%%%%%%%%%%%%%%%%%%%%%%%%%%%%%%%%%%%%%%%%%%%%%%%%%%%%%
\subsection{Anomalon masses and mixings}

Replacing the Higgs doublet by its VEV ($v_H \simeq 174$ GeV) gives mass mixing terms between the two electrically-charged anomalons, $E$ and $L^e$,
as well as between the two neutral anomalons,  $N$ and $L^\nu$. The mass matrix for the neutral anomalons takes the form
\be
{\cal L}_{N \rm mass} = - \left( \overline N_R  \; , \,  \overline L_R^\nu   \right) 
\left( \ba{ccccc}  y_N  \,  \langle \phi \rangle  &  y_{NL} \, v_H  \,   \\  [2mm]
y_{LN} e^{i \theta_N} \,  v_H  &  y_L  \,  \langle \phi \rangle  \,  \\  
 \ea \right)
\left( \ba{c} N_L \\  L_L^\nu  \ea \right) + {\rm H. c.} 
\ee
This can be diagonalzed by a $U(2)_L\times U(2)_R$ transformation. 
The charged anomalons $E$ and $L^e$ have an analogous mass matrix.

Let us restrict attention to the case where the off-diagonal masses are much smaller than the diagonal ones:
\be
\varepsilon_N \equiv \frac{\left( y_{LN}^2 + y_{NL}^2 \right) v_H^2 }{ \left( y_L - y_N \right)^2 \langle \phi \rangle^2 } \ll 1 ~~,
\ee
and the analogous condition for the charged anomalons, $\varepsilon_E \ll 1$.
This implies that the bulk of the anomalon masses preserves the electroweak symmetry, and the mixings are small. 
To leading order in $\varepsilon_N$ or $\varepsilon_E$, the physical masses are given by $m_{N_S} \simeq y_N \langle \phi \rangle$, 
$m_{E_S} \simeq y_E \langle \phi \rangle$, and $m_{N_D} \simeq m_{E_D} \simeq y_L \langle \phi \rangle$.
The mass degeneracy between the two states that are mostly part of the doublet is lifted by electroweak effects.
The mass ordering of these two states is important for phenomenology, so let's include the effects of order $\varepsilon_N$ or $\varepsilon_E$. 

For simplicity, let's assume 
\be
e^{i \theta_N}=e^{i \theta_E} = - 1 ~~,
\ee
as well as 
\be
 y_{LN} = y_{NL}  \;\; , \; \;  y_{LE} = y_{EL}    ~~.
 \label{particular}
\ee
Deviations from these assumptions are discussed later on.
The mass difference between the charged and neutral physical states that are mostly part of the weak-doublet anomalon is given by
\be
m_{E_D} - m_{N_D} \simeq  \frac{v_H^2 }{2  \langle \phi \rangle } 
 \left( \frac{  y_{LE}^2 }{ y_L + y_E} - \frac{  y_{LN}^2 }{ y_L + y_N} \right)   ~~.
 \label{eq:degeneracy}
\ee 
From this expression it is clear that $m_{E_D} > m_{N_D}$ for a range of Yukawa couplings,  and $m_{E_D} < m_{N_D}$ for a different range.

The left-handed neutral anomalons are given in the mass eigenstate basis  by 
\be
\left( \ba{c} N_{S_L} \\  N_{ D_L} \ea \right) = 
\left( \ba{ccccc}  c_N  & - s_N  \,   \\  
 s_N  &  c_N  \,  \\  
 \ea \right)
\left( \ba{c} N_L \\  L_L^\nu  \ea \right)  ~~,
\ee
while the right-handed ones are
\be
\left( \ba{c} N_{S_R} \\  N_{D_R} \ea \right) = 
\left( \ba{ccccc}  c'_N  & s'_N  \,   \\  
 - s'_N  &  c'_N  \,  \\  
 \ea \right)
\left( \ba{c} N_R \\  L_R^\nu  \ea \right)   ~~.
\ee
Here $s_N$ and $c_N$ are the sine and cosine of the mixing angle between left-handed neutral anomalons, and $s'_N$ and $c_N'$ 
are the analogous quantities for the right-handed anomalons. Similarly, the sines of the mixing angles between $E$ and $L^e$ are labelled by
 $s_E$ and $s'_E$. For the particular case in Eq.~(\ref{particular}),
 \be
 s_N = s'_N \simeq \frac{ y_{LN} \, v_H }{ m_{N_D} \! + m_{N_S}}   ~~,  
 \label{eq:sN}
 \ee
and an analogous expression holds for $s_E = s'_E$.

%%%%%%%%%%%%%%%%%%%%%%%%%%%%%%%%%%%%%%%%%%%%%%%%%%%%%%%%%%%%%%%%
\subsection{Anomalon decay modes} 

The $U(1)_B$ charges shown in Table 1 imply that there is no renormalizable interaction that would allow the lightest anomalon to decay. Note that 
higher-dimensional operators such as $(\bar N_R^c d_R )(\bar u_R^c d_R)$ or $(\bar E_R^c u_R )(\bar d_R^c u_R)$
 would allow the decay of an anomalon into three quarks, but their coefficients 
are expected to vanish unless additional fields with certain charges are present.
In the minimal anomalon model, where there are no additional fields or higher-dimensional operators, 
a $Z_3$ subgroup of $U(1)_B$ remains unbroken. All the anomalons carry charge +2 under this $Z_3$, and thus the 
lightest of them is stable. The best candidate for the lightest anomalon is thus $N_S$.
For a range of parameters, the relic abundance of $N_S$ makes it a viable dark matter candidate. 
By contrast, $N_D$ is mostly part of the doublet and has a large coupling to the $Z$ so that 
if it were the lightest anomalon, then its mass would have to be above a few TeV in order to evade the limits from direct detection experiments.

A relic abundance of $N_S$ smaller than the dark matter abundance is acceptable, provided there exists an additional dark matter component.
For a too large relic abundance of $N_S$, 
this scenario can be viable in the presence of an even highly suppressed decay width for $N_S$, for example via higher-dimensional operators.
The direct detection experiments still set upper limits on the small $s_N$ and $s_N'$ mixings (as the $N_S$ couplings to the $Z$ are proportional to these), 
as well as on the $Z'_B$ coupling-to-mass ratio, as a function of $m_{N_S}$. These limits become very weak for $m_{N_S} \lesssim 10$ GeV, due to the small 
nuclear recoil. 

The decays of the other three anomalon physical states depend on their mass ordering.
Let's focus on the following  ordering $m_{E_S} > m_{E_D}  > m_{N_D} > m_{N_S} $. 
The $N_D$ may decay into $N_S h^0 $ or $N_S Z$. A straightforward 
derivation gives the following decay widths for the parameter choice in Eq.~(\ref{particular}), up to corrections of order $\varepsilon_N^2$:
\bear
&& \Gamma (N_D \!\to N_S h^0 ) \simeq     \Gamma_{0N}
 \left(1- 2 M_h^2 \; \frac{m_{N_D}^2 \! + m_{N_S}^2 \! + m_{N_S} m_{N_D}}{ \left(m_{N_D}^2 \! - m_{N_S}^2 \right)^2} \right)   ~~,
\nonumber \\ [3mm]
&& 
\Gamma (N_D \!\to N_S Z ) \simeq  \Gamma_{0N}
\left(1- 6 M_Z^2  \frac{m_{N_S} m_{N_D}}{ \left(m_{N_D}^2 \! - m_{N_S}^2 \right)^2} \right)   ~~,
\label{eq:Nwidths}
\eear 
where the width for $M_h, M_Z \to 0$  is 
\be
\Gamma_{0N} = \frac{y_{LN}^2}{64\pi} \, m_{N_D} \left(1+ \frac{m_{N_S}}{m_{N_D}} \right)   \left(1- \frac{m_{N_S}}{m_{N_D}} \right)^{\! 3}  ~~.
\ee
In the $m_{N_D} - m_{N_S} \gg M_h$ limit,  $\Gamma (N_D \!\to N_S Z ) = \Gamma (N_D \!\to N_S h^0)$.
This is a consequence of the equivalence theorem: in the limit where the electroweak symmetry is unbroken, 
$N_D$ decays into $N_S$ plus a neutral component of the Higgs doublet $H$ through the Yukawa couplings of 
Eq.~(\ref{Higgs-Yukawa}). Thus, the $N_D \to N_S$ transitions have the width into $h^0$ equal to that into a longitudinal $Z$ up to corrections of order 
$(v_H/\langle \phi \rangle)^2.$ 

Note that a third decay mode of $N_D$ may also exist if the radial degree of freedom in $\phi$, labelled here by $\varphi$,
is lighter than  $m_{N_D} - m_{N_S}$. The $N_D \to N_S \varphi$ decay may be followed by $\phi \to WW$ or $ZZ$ if
there is $\varphi-h^0$ mixing. 
 
The $E_D^\pm$ anomalon has two decay modes: $N_D W$ and $N_S W$. 
The mass difference between the charged and neutral physical states that are mostly part of the weak-doublet anomalon is small, as shown in Eq.~(\ref{eq:degeneracy}).
This approximate mass degeneracy implies that the width of the $E_D^- \to N_D W^-$ and $E_D^+ \to \bar N_D  W^+$ decays 
is phase-space suppressed. In fact these are likely to be 3-body decays that proceed through an off-shell $W$ boson: $E_D \to  N_D W^* \to N_D  jj $ or $ N_D \ell \nu$. 
The $E_D \to N_S W$ decay width is typically not phase-space suppressed because $m_{E_D}$ may be substantially larger than $m_{N_S}+M_W$, but it is 
suppressed by $s_N^2$, the square of the small mixing given in Eq.~(\ref{eq:sN}). 
For $y_{LN} \ll y_{LE}$, the $B(E_D \to N_D W)$ branching fraction can be large (even for 3-body decays), 
as $m_{E_D} - m_{N_D}$ is not suppressed by $y_{LN}$ [see Eq.~(\ref{eq:degeneracy})].
For comparable values of $y_{LN}$ and $y_{LE}$, the 3-body decay has a tiny branching fraction, $B(E_D \to N_D W) \ll 1$.

Finally, the main decay modes of the heavier electrically-charged  anomalon, $E_S$, are $ N_D W$ or $E_D Z$ or $N_D  h^0$.
For $(v_H/\langle \phi \rangle)^2 \ll 1$, these three branching fractions are approximately in the 2:1:1 ratios \cite{Perelstein:2003wd} due to the equivalence theorem.
Note that the $E_S \to N_S W$ transition is suppressed by two more powers of the mixing angles.
An additional decay mode, $E_S \to E_D \varphi$, is relevant if the $\varphi$ scalar is light enough.

%%%%%%%%%%%%%%%%%%%%%%%%%%%%%%%%
\subsection{Signatures with a lepton pair, missing energy and jets}

$Z'_B$ production at the LHC can be followed by $Z'_B$ decay into a pair of anomalons. 
These would undergo cascade decays through the lighter anomalons, ending with a $N_S\bar N_S$  pair.
Given that the $Z_3$ subgroup of $U(1)_B$ keeps $N_S$ stable, the signatures involve missing transverse energy
and a few heavy SM bosons.

In the case of the  $m_{E_S} > m_{E_D}  > m_{N_D} > m_{N_S} $ ordering discussed above,
the $N_D$ neutral anomalon decays into $N_S Z$, due to the mixing between the two neutral gauge eigenstates ($N$ and $L^\nu$).
Pair production of  $N_D$  then leads to a $ZZN_S\bar N_S$ final state.
If the mass splitting between $N_D$ and $N_S$  is smaller than $M_Z$, then the decay proceeds through an off-shell $Z$: $N_D \to  N_S Z^* \to N_S j j $ or 
$N_S \ell^+\ell^- $ or $N_S \nu\bar\nu $. 
For $m_{N_D} - m_{N_S} > 125$ GeV,  $N_D$ can also decay into $N_S h^0$, so the $Zh^0N_S\bar N_S$ and $h^0h^0 N_S\bar  N_S$ final states are also possible.
The $B(N_D \to  N_S Z) \equiv B_{NZ}$ branching fraction depends on the $y_{NL}$ and $y_{LN}$ Yukawa couplings, 
on the $m_{N_D} - m_{N_S}$ mass difference, and on the mass of $\varphi$, $m_\varphi$. 
Assuming $m_\varphi > m_{N_D} - m_{N_S}$, which implies $B(N_D \to N_S h^0)  = 1-B_{NZ} $, the $B_{NZ}$ branching fraction is larger than $50\%$, as follows from Eq.~(\ref{eq:Nwidths}).

The  $E_D^\pm$ charged anomalon may decay into $N_D  W$, and into $N_S W$. As mentioned before, the former mode is 
likely to be a 3-body decay: $E_D  \to N_D  jj $ or $ N_D \ell \nu$, with 
$N_D$ subsequently decaying into $N_S$ and a $Z$ or $h^0$.
This decay mode, of branching fraction $B(E_D  \to N_D W) \equiv B_{NW}$, 
competes with the $E_D \to N_S W$ decay, which is suppressed by the small $N$-$L^\nu$ mass mixing.
 The $Z' \to E_D^+ E_D^-$ decay then leads to the $W^{(*)}W^{(*)} N_S\bar N_S + n (Z/h^0)$ final states, where the number of  $Z$ or Higgs bosons is $n = 0,1,2$.
 
For $M_\varphi >  m_{E_S} -  m_{E_D}$, the $E_S^\pm$ anomalon decays predominantly into $ N_D W$ or $ E_D Z$ or $E_D h^0 $.
Given that $m_{N_D} + M_W < m_{E_D} + M_Z$,  its branching fractions satisfy $ B(E_S \to N_D W)  \equiv B_{EW}  > 50\%$, 
and  $B(E_S \to E_D h^0 ) < B(E_S \to E_D Z) \equiv B_{EZ} < 25\%$. 
 The $Z' \to E_S^+ E_S^-$ decay gives the $W^+W^-N_S\bar N_S+ n (Z/h^0)$ final states with $n = 2,3,4$. 
 
The leptonic decays of the SM bosons produced in the above cascade decays provide various opportunities for probing the leptophobic $Z'$.
An example is the signal that includes $Z\to \ell^+\ell^-$, large $\met$, and two or more jets arising from the decays of the other SM bosons produced in the 
cascade decays. The ATLAS collaboration has reported a 3$\sigma$ excess in this channel \cite{Aad:2015wqa}, 
based on the observation of 29 events for a background of $10.6 \pm 3.2$ events.
The CMS search \cite{Khachatryan:2015lwa} in the same channel is consistent with the SM prediction, but the event selection is less stringent compared to the 
ATLAS search, so that the background is about 7 times larger; in particular, the ATLAS search requires $H_T > 600$ GeV, a condition easily satisfied 
by events associated with the decay of a $Z'$ boson of mass above $\sim 1$ TeV.

\begin{figure}[t]
\begin{center} 
{
\unitlength=2. pt
\SetScale{1.05}
\SetWidth{1.}      % line    size control
\normalsize    %  letter  size control
{} \allowbreak
\begin{picture}(100,100)(7,-37)
\ArrowLine(4,80)(22,50)
\ArrowLine(22,50)(4,20)
\Photon(22,50)(65,50){3}{6}
\ArrowLine(65,50)(85,75)\Photon(85,75)(112,104) {1.2}{5} \ArrowLine(85,75)(132,84)  
\ArrowLine(85,25)(65,50)\DashLine(112,-4)(85,25){3}\ArrowLine(132,14)(85,25)
\Text(-1,39)[c]{$q$}\Text(-1,14)[c]{$\bar q$}\Text(23,33)[c]{\small $ Z'$}
\Text(36,38)[c]{\small $N_D$}\Text(36,13)[c]{\small $\overline N_D$}
\Text(76,43)[c]{\small $N_S$}\Text(76,6)[c]{\small $\overline N_S$}
\Text(63,55)[c]{\small $Z$}\Text(64,-1)[c]{\small $h^0$}
\end{picture}
\begin{picture}(100,100)(8,-37)
\ArrowLine(4,80)(22,50)
\ArrowLine(22,50)(4,20)
\Photon(22,50)(65,50){3}{6}
\ArrowLine(65,50)(85,75)
\ArrowLine(85,25)(65,50)
\Photon(85,75)(112,104) {1.2}{5} \ArrowLine(85,75)(137,84) \Photon(137,84)(172,105){1.2}{5}  \ArrowLine(137,84)(197,86)  
\DashLine(85,25)(114,-5){3} \ArrowLine(120,16)(85,25)\Photon(120,16)(152,-5){1.2}{5}  \ArrowLine(160,11)(120,16) \Photon(190,-4)(160,11){1.2}{5} \ArrowLine(205,11)(160,11)  
\Text(-1,39)[c]{$q$}\Text(-1,14)[c]{$\bar q$}\Text(23,33)[c]{\small $ Z'$}
\Text(36,38)[c]{\small $E_S^+$}\Text(36,13)[c]{\small $E_S^-$}
\Text(64,58)[c]{\small $W^+$}\Text(110,45)[c]{\small $N_S$} \Text(95,58)[c]{\small $Z$}
\Text(65,-2)[c]{\small $h^0$}\Text(87,-3)[c]{\small $W^-$}\Text(115,6)[c]{\small $\overline N_S$} \Text(104,-2)[c]{\small $Z$}
\Text(64,37)[c]{\small $N_D$} \Text(54,17)[c]{\small $E_D^-$} \Text(74,12)[c]{\small $\overline N_D$}
\end{picture}
}
\end{center}
\vspace*{-2.8cm}
\caption{Representative processes for cascade decays of $Z'_B$ via anomalons. The final states include $ \slash \!\!\! \!  E_T$ 
due to the $N_S$ anomalons, two or more jets from decays of some SM bosons, and leptons, for example from $Z\to \ell^+\ell^-$.}
\label{fig:ZN}
\end{figure}
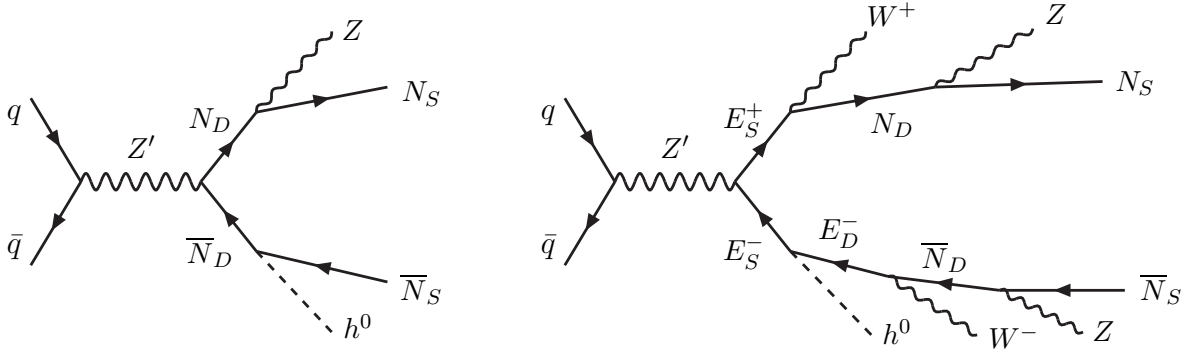

Let us identify a region of parameter space in the $Z'_B$ model that could produce the 18 or so excess events observed by ATLAS (different explanations  are discussed in \cite{Barenboim:2015afa}).
The combined branching fraction for this signal arising from $Z' \to N_D\bar N_D$ is
\bear
&& \hspace*{-0.81cm}   
B(Z' \to N_D\bar N_D \to (\ell^+\ell^-) jj  \met X) =
2B(Z' \to N_D\bar N_D) \, B_{NZ} \, B(Z \to \ell^+\ell^-)   
 \nonumber \\ [2mm]
&&  \hspace*{1.4cm}    \times 
 \left[  (1-B_{NZ})\,  B(h^0 \to jjX )    +  B_{NZ} \, B(Z \to jjX)  \right]  ~,
\eear
where the lepton pair labelled by $(\ell^+\ell^-) $ is produced by a $Z$ decay. The first term in the above square bracket arises from the 
processes shown in the left diagram of Figure \ref{fig:ZN}. The other term arises from a similar diagram replaced by a $Z$ decaying hadronically.
The total branching fraction for the inclusive Higgs boson decays into two jets and anything else is 
\bear
&&  \hspace*{-0.7cm}  
B(h^0   \!\!\to \! jjX )  = B(h^0  \!\!\to \! b\bar b)  + B(h^0   \!\!\to \! gg) + B(h^0   \!\!\to \! c\bar c ) 
 + B(h^0   \!\!\to \! \tau \tau) B_{\rm had}(\tau)^2    
 \nonumber \\ [2mm]
&&  \hspace*{1.5cm}  
+ B(h^0 \to WW^* \to jjX ) + B(h^0 \to ZZ^* \to jjX )   
 \nonumber \\ [2mm]
&&  \hspace*{1.5cm}  
\approx 93.6\%
\label{HiggsB}
\eear
The SM predictions for the Higgs branching fractions \cite{Heinemeyer:2013tqa} at $M_h = 125 $ GeV are 
57.7\%, 8.6\%, 2.9\%, 6.3\%, 19.3\%, 2.5\%, respectively, for the six decay modes shown on the right-hand side of the above equation.
The decays via $WW^*$ and $ZZ^*$ include contributions from hadronic tau decays (which appear in the detector as jets, if 
 tau identification methods are not deployed), and $B_{\rm had}(\tau) \equiv B(\tau \to {\rm hadrons} +\nu ) \approx 64.8\%$. 
 Similarly to Eq.~(\ref{HiggsB}), $B(Z \to jjX) \approx 71.3\%$.

The same leptons-plus-jets-plus-$\met$ signal arising from $Z' \to E_D^+\bar E_D^-$ has a combined branching fraction of
\be
B(Z'\! \to \!  E_D^+ E_D^- \!\to\! (\ell^+\ell^-) jj  \met X)  \approx  2 B(Z' \!\to\! E_D^+  E_D^- ) \, B_{NW} \, B_{NZ} B(Z\! \to\! \ell^+\ell^-)  ~.
\ee
Here a negligible term (corresponding to the case where none of the SM bosons decays into jets) was included for simplicity. 
Finally, the combined branching fraction of the contribution from $Z' \to E_S^+\bar E_S^-$ is
\bear
&& \hspace*{-2.5cm}   
B(Z' \to E_S^+  E_S^- \to (\ell^+\ell^-) jj  \met X) \approx 2 B(Z' \to E_S^+  E_S^-) \,  B(Z \to \ell^+\ell^-) \,  
 \nonumber \\ [2mm]
&&  \hspace*{1.1cm}   
\times \left[ B_{EZ} + B_{EW} \, B_{NZ}   + \left(  1 -  B_{EW}  \right) \, B_{NW}  \, B_{NZ}   \right]  ~.
\eear
One of the cascade decays contributing here is shown in the right diagram of Figure \ref{fig:ZN}.

Let us consider a benchmark case: $M_{Z'} = 1.5$ TeV, $m_{N_S} = 10$ GeV,  $m_{N_D} = 400$ GeV, $m_{E_D} = 420$ GeV, $m_{E_S} = 600$ GeV, 
and $M_\varphi  > m_{E_S}$.
In this case $B_{NZ} \approx 56\%$, $B_{EZ} \approx 1/4$, $B_{EW} \approx 1/2$, while $B(E_D \to N_D W) = B_{NW}$ can be treated as a free parameter. The 
relevant $Z'$ branching fractions are
$B(Z' \to N_D^+  N_D^-) \approx  20.8\%$,  $B(Z' \to E_D^+  E_D^-) \approx 20.3\%$,  and $B(Z' \to E_S^+  E_S^-) \approx  14.7\%$.
This gives the following total branching fraction for the signal
\be
B(Z' \to  (\ell^+\ell^-)  jj  \met X) \approx  2.3\% + 2.1\% \, B_{NW}  ~~,
\ee
with $Z' \to N_D \bar N_D$  contributing $1.3\%$. 
Multiplying the above total branching fraction of roughly 4\% with a production cross section of 70 fb (corresponding for example to  $M_{Z'} = 1.5$ TeV and $g_z \approx 0.5$)
and with an acceptance times efficiency of the event selection of about 30\%,
gives 17 events in 20 fb$^{-1}$ of data. 

An interesting feature of the ATLAS excess is that it includes 9 events with 5 or more jets, with the background for these being only 1.3 events (see Figure 7 of \cite{Aad:2015wqa}). 
This is consistent with the longer cascade decays arising from $Z' \to E_S^+  E_S^-$ and $Z' \to E_D^+  E_D^-$, which include jets from the decays of two or more 
SM bosons. 

The rate for $Z'_B$ production increases by a factor of 4.6 (6.5) for $M_{Z'} = 1.5 $ TeV ($M_{Z'} = 1.8 $ TeV)
at $\sqrt{s} = 13$ TeV compared to $\sqrt{s} = 8$ TeV, so the hypothesis that the ATLAS excess is due to $Z'_B$ 
cascade decays will be tested in Run II of the LHC.

Another signal that includes leptons is $Z' \to  (\ell^+\ell^-) \ell  jj  \met$ where the  third lepton arises from a $W$ decay. 
The branching fraction is smaller in this case by a factor of approximately 5.

The jets plus $\met$ final state is also potentially interesting in this model. The combined branching fraction for $Z' \to N_D\bar N_D \to  4j+\met$ is 10\%,
so there would be on the order of 40 events produced in Run I for the benchmark case with $M_{Z'} = 1.5$ TeV. 
This is slightly smaller than the uncertainty in the background reported in Table 1 of \cite{Chatrchyan:2014lfa}, so this channel is also promising in Run II.
Similarly, the combined branching fraction for $Z'$ to six or more jets and $\met$ is $\sim 4\%$, which gives a number of signal events comparable
to the current uncertainty in the background. 

\medskip
  
 %%%%%%%%%%%%%%%%%%%%%%%%%%%%%%%%%%%%%%%%%%%%%%%%%%%%%%%%%%%%%%%%%%%%%%%%%%%
%%%%%%%%%%%%%%%%%%%%%%%%%%%%%%%%%%%%%%%%%%%%%%%%%%%%%%%%%%%%%%%%%%%%%%%%%%%
\section{$U(1)_{R12}$  model}
\setcounter{equation}{0}

In the $Z'_{R12}$ model, introduced in Section 2.2, the only SM fields carrying the $U(1)_{R12}$ gauge charge
are the right-handed quarks of the first and second generations. The masses of the $c$ and $u$ quarks arise from dimension-5 operators of the type
$\bar c_R Q_L H \phi$, while the $s$ and $d$ quark masses arise from similar operators with  $\phi$ replaced by  $\phi^\dagger$.

The $[U(1)_Y]^2 U(1)_{R12}$ and $U(1)_Y [U(1)_{R12}]^2$ gauge anomalies can be cancelled by the inclusion of two charged anomalons, $E$ and $E'$
which transform under the SM gauge group the same way as the weak-singlet leptons. Imposing that these
get masses from Yukawa couplings to $\phi$, and that at least one of them mixes with SM leptons so that it can decay\footnote{An alternative is that the charged anomalons decay via higher-dimensional operators, in which case they can all have nonzero  $U(1)_{R12}$ charges. Anomaly cancellation solutions of this type 
exist in the presence of at least two SM-singlet anomalons.}, 
gives the  $U(1)_{R12}$ chiral charges shown in Table 2.
The remaining $[U(1)_{R12}]^3$ and $U(1)_{R12}$-gravitational anomalies are cancelled by a single SM-singlet Weyl fermion, $N_R$, of $U(1)_{R12}$ charge +2.

The neutral anomalon, $N_R$, remains massless unless certain higher-dimension operators are introduced (such as $(\phi^\dagger)^4 \overline N^c_R N_R$ or $(\phi^\dagger)^2 \overline L_L H N_R$). 
A massless $N_R$ is not problematic though: it interacts only through $Z'_{R12}$ exchange, so that for $M_{Z'} \sim O(1$ TeV)  it decouples 
in the early universe well before primordial nucleosynthesis, and thus it contributes significantly less than an additional neutrino species.
% phenomenologically

%%%%%%%%%%%
\begin{table}[t!]\begin{center}
\renewcommand{\arraystretch}{1.2}
\begin{tabular}{cc|ccc|c}\hline 
field  & spin & $SU(3)_c$   & $SU(2)_W$   &  $U(1)_Y$  & $U(1)_{R12}$ \\ \hline\hline
$u_R$ , $c_R$    &  &    &  & +2/3  & +1   \\ [-.6em]
          & 1/2      & \ 3 \  & \  1 \ &  &    \\ [-.6em]
$d_R$ , $s_R$    &   &   & & $-1$  & $-1$    \\ [0.2em]  \hline
$E_L$ , $E'_L$  & & &        &                   &  \hspace*{5mm}  $\! +1$ ,  $ -1$ \hspace*{2mm}  \\ [-.6em]
              & 1/2    & \ 1 \   & \  1 \ & \  $-1$  &    \\ [-.6em]
$E_R$ , $E'_R$ &  &  &     &                     &  \hspace*{8mm}  $\! 0$ \hspace*{0.1mm}  ,  $ -2$  \hspace*{3mm}  \\ [0.1em]
 \hline
 $N_R$   & 1/2      & \ 1 \  & \  1 \ & \  $0$ \ &  $\! +2$  \\ [0.1em]
  \hline
$\phi$ & 0 & \ 1 \  & \  1 \ & \  0 \  &   \hspace*{2mm}  $\!  +1$  \hspace*{2mm}  \\ \hline
\end{tabular}
\medskip \\
\vspace{0.2in}
\caption{The $U(1)_{R12}$-charged SM quarks and the fields beyond the SM 
in the minimal $Z'_{R12}$ model. 
All the anomalons ($E$, $E'$, $N_R$) are weak- and color-singlets.}
\label{table:U1R}
\end{center}
\end{table}
%%%%%%%%%%%

%%%%%%%%%%%%%%%%%%%%%%%%%%%%%%%%
\subsection{Properties of the charged anomalons}

The Yukawa couplings involving two charged anomalons are given by
\be
- y_E \, \phi \, \overline E_L E_R  - y'_E \, \phi \, \overline E'_L E'_R  -  y^{\prime\prime}_{E} \, \phi^\dagger \overline E'_L E_R + {\rm H.c.}
\ee
These induce mass terms for $E$ and $E'$, as well as a mass mixing between these fermions. After the diagonalization of their $2\times 2 $
mass matrix, the physical states $E_1$ and $E_2$ (of masses $m_{E_1} < m_{E_2}$) 
have both diagonal and  off-diagonal couplings to $Z'_{R12}$.
Let us express the relations between mass and gauge eigenstates as
\bear
& E_{1L} = E'_L c_{\theta_L} + E_L s_{\theta_L}  ~~,    &  \hspace*{1.3cm}   E_{1R} = - E'_R c_{\theta_R} + E_R s_{\theta_R}  ~~,   
\nonumber \\
& E_{2L} = - E'_L s_{\theta_L} + E_L c_{\theta_L} ~~,   &  \hspace*{1.3cm}   E_{2R} = E'_R s_{\theta_R} + E_R c_{\theta_R} ~~,
\eear
where $s_{\theta_{L,R}} \equiv   \sin \theta_{L,R}$,  
$c_{\theta_{L,R}} \equiv \cos \theta_{L,R}$, and the left- and right-handed mixing angles ($\theta_L$ and $\theta_R$) are functions of the $y_E$, $ y'_E$ and $y^{\prime\prime}_{E}$ 
Yukawa couplings, which  without loss of generality can be taken to be real parameters. 
The two mixing angles are related by
\be
\tan  \theta_R =  \frac{m_{E_1}}{m_{E_2}} \tan \theta_L    ~~.
\ee
The couplings of the mass-eigenstate anomalons to $Z'_{R12}$ are
\bear
\frac{g_z}{2} Z'_{R12 \, \mu} &&  \hspace*{-0.7cm}  
\left[  - 2 \left( c_{\theta_R}^2 \overline E_{1_R} \gamma^\mu E_{1_R} +  s_{\theta_R}^2 \overline E_{2_R} \gamma^\mu E_{2_R} \right)
+ (c_L^2-s_L^2) \left( \overline E_{1_L} \gamma^\mu E_{1_L} - \overline E_{2_L} \gamma^\mu E_{2_L} \right) \right.
\nonumber \\
&& + \, \left. 2 \left(  s_{\theta R} c_{\theta R}   \overline E_{1_R} \gamma^\mu E_{2_R} + s_{\theta L}c_{\theta L}   \overline E_{1_L} \gamma^\mu E_{2_L} + {\rm H.c.} \right) 
\right] ~~.
\label{Z'R12-mass-eigenstates}
\eear

The branching fractions of $Z'_{R12}$ for $M_{Z'} \gg 2 m_{E_2}$ are given by $B (Z'_{R12} \to jj) = 6/11$, $B (Z'_{R12} \to N_R \bar N_R) = 2/11$, 
and
\bear
&& B (Z'_{R12} \to E_1^+ E_1^-) = \frac{1}{22} \left[  (c_{\theta_L}^2-s_{\theta_L}^2)^2 + 4  c_{\theta R}^4 \right]   ~~,
\nonumber \\ [2mm]
&& B (Z'_{R12} \to E_2^+ E_2^-) =  \frac{1}{22} \left[  (c_{\theta_L}^2-s_{\theta_L}^2)^2 + 4  s_{\theta R}^4 \right]  ~~,
\nonumber \\ [2mm]
&& B (Z'_{R12} \to E_1 E_2) =  \frac{4}{11} \left(  c_{\theta_L}^2s_{\theta_L}^2 + c_{\theta R}^2 s_{\theta R}^2 \right)  ~~,
\label{BRE1E2}
\eear
The total $Z'_{R12}$ width is 
\be
\Gamma (Z'_{R12}) = \frac{11g_z^2}{48\pi} M_{Z'}  ~~,
\ee
implying that the maximum value for $g_z$ that keeps the width below $5\% M_{Z'}$ is 
\be
g_z^{\rm max} = (12\pi/55)^{1/2} \approx 0.83  ~~.
\label{gzmax}
\ee
For this value of $g_z$, the rate for $Z'_{R12}$ production is 3.1 times larger than the cross section shown in Figure~1.

Note that the sum of the branching fractions (\ref{BRE1E2}) into electrically-charged anomalons  is 3/11.
The asymmetric $Z'_{R12} $ decays, into $E_1^+ E_2^-$ or  $E_1^- E_2^+$, are particularly interesting for phenomenology.
Their branching fraction can be as large as 2/11.

There are also Yukawa couplings involving a SM lepton and an anomalon:
\be
- y_{Ee}  \phi^\dagger  \overline E'_L e_R^j  - y'_{Ee}  \phi \overline E_L e_R^j -  y_{EL}  H \overline L_L^j  E_R   + {\rm H.c.} 
\label{mixedYukawa}
\ee
These induce mass mixings between the electrically-charged  SM leptons and the anomalons. As a result, the $E_1$ and $E_2$ can decay into a SM lepton and
a SM boson. 
It is technically natural to assume that the above couplings are very small, in which case they will not affect the masses and mixings of $E_1$ and $E_2$ but they would allow their  decays. There are several decay modes: $E_{1,2} \to W \nu, \, Z\ell , \, h^0 \ell$, where $\ell = e,\mu,\tau$.
Given that the Yukawa couplings do not have to be equal for different lepton flavors, the $E_1$ and $E_2$ decays are likely to violate lepton universality, {\it e.g.} 
$B( E_1 \to  Z e) \neq B( E_1 \to  Z \mu)$. For $m_{E_1} \gg M_h + m_\tau$, using the equivalence theorem 
and summing over the three lepton flavors
gives
\bear
&& B(E_1 \to W \nu) \approx 50\%   ~~,
\nonumber \\
&& B(E_1 \to Z e ) + B(E_1 \to Z \mu) + B(E_1 \to Z \tau)  \approx 25\%  ~~,
\nonumber \\
&& B(E_1 \to h^0 e ) + B(E_1 \to h^0 \mu) + B(E_1 \to h^0 \tau)  \approx 25\%    ~~.
\label{E1decays}
\eear
The $E_2$ decays into SM states  compete with the $E_2 \to E_1 Z^{\prime *}_{R12}$ decays induced by the off-diagonal couplings in Eq.~(\ref{Z'R12-mass-eigenstates}), where the off-shell $Z^{\prime *}_{R12}$ then goes into $N\bar N$
or $jj$, or even $E_1^+ E_1^-$ if $m_{E_2} > 3 m_{E_1}$. The ratio of the $E_2 \to E_1  N\bar N$ and $E_2 \to E_1 jj$ 
branching fractions is 1/3. Whether these branching fractions are negligible or dominate over the $E_2 \to W \nu, \, Z\ell , \, h^0 \ell$ decays 
depends on the Yukawa couplings shown in Eq.~(\ref{mixedYukawa}), on the value of $g_z$, and on the $E_2-E_1$ mass splitting.

%%%%%%%%%%%%%%%%%%%%%%%%%%%%%%%%
\subsection{Non-universal signatures with leptons and jets}

The production of $Z'_{R12}$ in the $s$ channel at the LHC would be followed by several possible cascade decays that lead to interesting final states
with leptons, jets and possibly missing energy. 
The $Z'_{R12} \to E_1^+ E_1^- $ process followed by the decays shown in Eq.~(\ref{E1decays}) produces final states with missing energy,   
\be
Z'_{R12}  \to E_1 ^+ E_1^- \to  \; W^+ \bar\nu \, W^- \nu \;\; ,  \;\; W \nu \, Z \ell  \;\; ,  \;\; W \nu \, h^0 \ell    ~~,
\label{Wnu}
\ee
or final states with one or more pairs of leptons,  
\be
Z'_{R12}  \to E_1 ^+ E_1^- \to  \; h^0 \ell \, Z \ell' \;\; ,  \;\; h^0 \ell \, h^0 \ell'  \;\; ,  \;\;  Z \ell \, Z \ell'  ~~.
\ee
The leptons denoted here by $\ell$ and $\ell'$ may each be an $e$, a $\mu$ or a $\tau$, with branching fractions 
that are expected to violate lepton universality.

Let us assume that the leptons are predominantly electrons. The hadronic 
decays of the SM bosons then lead to the $e^+e^- \! +$jets or $e\nu +$jets final states with total branching fractions of 
$B(Z'_{R12}  \! \to\! E_1 ^+ E_1^- ) $ times a factor of 
$[B_{\rm had}(Z) + B_{\rm had}(h^0)]^2/16$ or $B_{\rm had}(W)  [B_{\rm had}(Z) + B_{\rm had}(h^0)]/4$, respectively.
Here $B_{\rm had}(Z)$ is the sum of the branching fractions of the $Z$ decays into hadrons that appear as jets. For example,
$Z\to \tau^+\tau^-$ should be included when the taus decay hadronically, if the event selection does not 
explicitly identifies hadronic tau events. Similarly, $B_{\rm had}(h^0)$ includes, for example, decays  into $WW^*$ followed
by hadronic $W$ decays.

An interesting case is that where $m_{E_1} \gg M_h$, so that the SM bosons are highly boosted. The jets arising from a boson decay are then
collinear, and appear as a single jet, labelled by $J$. For not so large values of  the $m_{E_1}/M_h$ ratio, $J$ may be broader than usual QCD jets.
The signatures of the $Z'_{R12}  \! \to\! E_1 ^+ E_1^-$ processes discussed above are thus $e^+e^- JJ$ and $e\nu JJ$. 
These provide an alternative interpretation of the excess events reported by the CMS collaboration in the $W_R \to e^+e^-jj$ 
search \cite{Khachatryan:2014dka} as well as in the $(e^+j)(e^-j)$ and $(ej)(\nu j)$ ``first generation leptoquark" searches \cite{CMS:2014qpa}

This interpretation of the CMS excess electron events in terms of a $Z'$ decaying via SM-vectorlike leptons has some common features with that given
in Ref.~\cite{Dobrescu:2014esa}. An important difference is that the SM-vectorlike leptons in the present case have chiral couplings to the $Z'$.
Furthermore, their existence is required by the anomaly cancellation conditions. Another difference is that the $Z'$ decays in the models 
presented in \cite{Dobrescu:2014esa} involved additional jets.
Other interpretations of the CMS excess electron events have been proposed in \cite{Bai:2014xba}.

In the case of a boosted $Z$ or $W$,  the decays that involve hadrons have branching fractions
\bear
&& B_{\rm had}( Z) = B( Z \to jj) + B(Z \to \tau^+\tau^-) B_{\rm had}(\tau) (2- B_{\rm had}( \tau) ) \approx 72.9\% ~~,
\nonumber \\ [3mm]
&& B_{\rm had}( W ) = B( W\! \to jj) + B(W \!\to \tau \nu ) B_{\rm had}( \tau) \approx 74.4\% ~~.
\eear
Similarly, for a boosted Higgs boson, any of its decays that involve hadrons would appear as a single jet, so
\bear
B_{\rm had}( h^0 )  \hspace*{-0.21cm}  &=&   \hspace*{-0.21cm} 
B( h^0\!  \to b\bar b)+ B( h^0 \! \to gg) + B( h^0 \! \to c\bar c)  + B(h^0 \! \to \tau^+\tau^-) B_\tau(2- B_\tau) 
\nonumber \\ 
&+&   \hspace*{-0.21cm}  B( h^0\! \to WW^*) B_{\rm had}(W) \left[2- B_{\rm had}(W) \right] +  B( h^0\! \to ZZ^*) B_{\rm had}(Z ) \left[2- B_{\rm had}(Z) \right] 
\nonumber \\ [1mm]
& \approx &  \hspace*{-0.21cm}  97.2\% ~~, 
\eear 

\begin{figure}[t]
\begin{center} 
{
\unitlength=2. pt
\SetScale{1.05}
\SetWidth{1.}      % line    size control
\normalsize    %  letter  size control
{} \allowbreak
\vspace*{-0.3cm}
\begin{picture}(100,100)(-10,-37)
\ArrowLine(4,80)(22,50)
\ArrowLine(22,50)(4,20)
\Photon(22,50)(65,50){3}{6}
\ArrowLine(65,50)(85,75)\Photon(85,75)(112,104) {1.2}{5} \ArrowLine(85,75)(132,84)  
\ArrowLine(85,25)(65,50)\DashLine(112,-4)(85,25){3}\ArrowLine(132,14)(85,25)
\Text(-1,39)[c]{$q$}\Text(-1,14)[c]{$\bar q$}\Text(23,33)[c]{\small $ Z'$}
\Text(36,38)[c]{\small $E_1^+$}\Text(36,13)[c]{\small $E_2^-$}
\Text(76,43)[c]{\small $\ell^+$}\Text(76,6)[c]{\small $\ell^-$}
\Text(63,55)[c]{\small $Z$}\Text(64,-1)[c]{\small $h^0$}
\end{picture}
\begin{picture}(100,100)(-18,-37)
\ArrowLine(4,80)(22,50)
\ArrowLine(22,50)(4,20)
\Photon(22,50)(65,50){3}{6}
\ArrowLine(65,50)(85,75)\Photon(85,75)(112,104) {1.2}{5} \ArrowLine(85,75)(132,84)  
\ArrowLine(85,25)(65,50)\DashLine(112,-4)(85,25){3}\ArrowLine(132,14)(85,25)
\Text(-1,39)[c]{$q$}\Text(-1,14)[c]{$\bar q$}\Text(23,33)[c]{\small $ Z'$}
\Text(36,38)[c]{\small $E_2^+$}\Text(36,13)[c]{\small $E_2^-$}
\Text(74,43)[c]{\small $\nu$}\Text(76,6)[c]{\small $\ell^-$}
\Text(64,55)[c]{\small $W$}\Text(64,-1)[c]{\small $h^0$}
\end{picture}
}
\end{center}
\vspace*{-2.8cm}
\caption{Representative processes for $Z'_{R12}$ decays into anomalons. The final states include, for example, $\ell^+\ell^- JJ$ (left diagram) or
$\ell \nu JJ$ (right diagram), where $J$ stands for the wide jet from the boosted hadronic decays of a SM boson.}
\label{fig:eejj}
\end{figure}
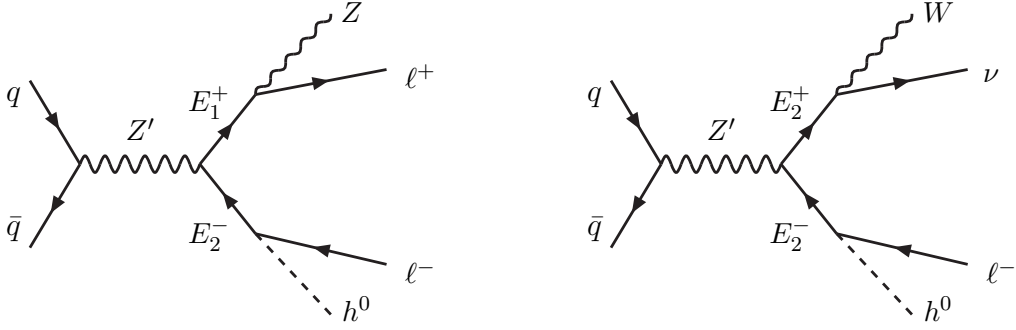

The $Z'_{R12} \to E_1E_2 $ and $Z'_{R12} \to E_2^+ E_2^- $ processes,  followed by the decay of $E_2$ into $W\nu$, $Ze$ or $h^0 e$
further contribute to the $e^+e^-\!+$jets and $e\nu +$jets final states (see Figure~\ref{fig:eejj}). Given that these contributions may be comparable to those from 
 $E_1 ^+ E_1^-$ production, it is expected that the $e J$  invariant mass distribution does not exhibit a clear peak near $m_{E_1}$, also in agreement with 
 the CMS observations \cite{CMS:2014qpa}. 
The total branching fractions of these signals are
\bear
&& B(Z'_{R12} \to e^+e^- + {\rm jets} ) \approx \frac{3}{176}   [B_{\rm had}(Z) + B_{\rm had}(h^0)]^2 \approx 4.9\%  ~~,
\nonumber \\ [2mm]
&& B(Z'_{R12} \to  e + \met + {\rm jets}) \approx \frac{3}{44} B_{\rm had}(W)  [B_{\rm had}(Z) + B_{\rm had}(h^0)] \approx  8.6\% ~~.
\eear
where the $E_2$ decays through an off-shell $Z'_{R12}$ have been assumed negligible.
For $M_{Z'} = 2$ TeV and maximal $g_z$ as in Eq.~(\ref{gzmax}), the leading-order $Z'_{R12}$ production cross section at the 8 TeV LHC is
approximately 140 fb. Thus, the rate for the $(J e)(J e)$ signal can be as large as 7 fb. With 20 fb$^{-1}$ of data, and an event selection efficiency 
of the order of 20\% percent it is easy to obtain the 10 signal events reported by CMS in the $W_R$ search, if $g_z \approx 0.5$.
The same set of parameters can also give 10 or so $(J e)(J \nu)$ signal events, which may explain the excess reported in the first-generation leptoquark search.

For a fraction $f_1$ of the 
 $(J e)(J e)$ signal events only one $J$ is formed by the  merger of a $b$ jet and a $\bar{b}$ jet, while for another fraction $f_2$ each $J$ is a merged $b\bar{b}$ jet.
Using the notation $B_b = [B(h^0 \to b\bar b) +  B(Z \to b\bar b)]/2 \approx 36.5\%$, the fractions of events are
 $f_1 = 2 B_b(1-B_b) \approx 45\% $  and $f_2 = B_b^2 \approx 15\%$. 
 In the case of the $(Je)(J\nu)$ events, a fraction $B_b$ of events include a single  $J =b\bar{b}$, and no events include two such jets.
 Detailed experimental information and further studies are necessary to
 assess whether these fractions are small enough to satisfy the CMS observation that the ``leptoquark"  $eejj$ and $e\nu jj$ events (including the background) 
 ``are not characterized by the presence of $b$ quarks" \cite{CMS:2014qpa}.

%%%

Processes of this type where the boosted Higgs boson decays to $b \bar b$ allow a reduction of the background by $b$ tagging the nearly collinear 
$b \bar b$ pair. Consider the $Z'_{R12}  \to E_i ^+ E_{i'}^- \to  \; h^0 e W\nu$ processes ($i,i' =1,2$), which have a combined 
branching fraction of $3/176$. 
Multiplying by the $h^0  \to b \bar b$ branching fraction gives $B(Z'_{R12}  \to (b\bar{b}) e W \nu ) \approx 0.91\%$. 
This leads to roughly two events for the parameters that may account for the $ eejj $ and $e\nu jj$ events discussed above.  
To avoid a further reduction of this small signal, it is preferable to consider the hadronic $W$ decays. The signal thus includes 
a $b\bar{b}$ pair with invariant mass near 125 GeV, an electron, missing energy, and an additional wide jet arising from the boosted $W$. 
Ignoring the latter, the background can be estimated from the CMS search for a $h^0 W \to (b \bar b)e\nu $ resonance \cite{Khachatryan:2014dka}
(although in the case of $Z'_{R12} $ the electron and the missing energy would not necessarily satisfy the $W$ mass constraint). 
Events with $(b\bar{b}) e \nu$ invariant mass in the $\sim 1.3-1.7$ TeV range are most relevant, as the 
hadronically decaying $W$ carries about 1/4 of the energy released in the $Z'_{R12} $ decay.
Figure 4 of Ref.~\cite{Khachatryan:2014dka} shows a background of about 1.2 events in that range.
Therefore, the signal for $M_{Z'} = 2$ TeV  looks too small given this background in the Run I data, but can be tested 
in the 13 TeV run of the LHC.

%%%%%%%%%%%%%%%%%%%%%%%%%%%%%%%%%%%%%%%%%%%%%%%%%%%%%%%%%%%%%%% 
%%%%%%%%%%%%%%%%%%%%%%%%%%%%%%%%%%%%%%%%%%%%%%%%%%%%%%%%%%%%%%%
\setcounter{equation}{0}
\section{\label{conclusions} Conclusions}

Extensions of the SM  gauge group typically require new fermions in order to cancel the anomalies. These anomalons have masses
that are usually comparable or smaller than those of the new gauge bosons. When the anomalons are lighter than half the gauge boson mass,
the decays of the new gauge bosons into anomalons may provide the discovery signatures at colliders.

In this paper I have shown that leptophobic $Z'$ bosons, even though they don't couple to leptons, may lead to interesting collider 
signatures involving leptons. These arise from cascade decays through anomalons, which lead to final states involving $Z$,  Higgs and $W$ bosons, plus
either stable neutral anomalons or SM fermions.
The former case occurs, for example, in the baryonic $U(1)_B$ model, were the minimal anomalon content (see Table 1) implies that a $Z_3$  
symmetry keeps the lightest anomalon stable. One of the ensuing signature is a leptonically decaying $Z$ plus $\met$ plus two or more jets, which 
may explain the ATLAS excess \cite{Aad:2015wqa} reported in this channel.

Another leptophobic $Z'$ model is that where only the weak-singlet quarks of the first and second generations carry the new charges. This $U(1)_{R12}$
model requires a slightly smaller anomalon content (Table 2), and its collider signatures violate lepton flavor. 
A few CMS excesses in channels involving electrons and jets may be accommodated in this model.

Independent of whether the small deviations from the SM predictions observed in the Run I at the LHC will be confirmed in Run II, resonant signatures involving 
leptons and jets as discussed here are an important test of extended gauge groups.

\bigskip

{\bf Acknowledgments:} \ I would like to thank Patrick Fox and Ann Nelson for insightful conversations.

%%%%%%%%%%%%%%%%%%%%%%

\end{document}